\documentclass[reprint,amsmath,amssymb,aps,prl]{revtex4-2}
\usepackage{amsthm}
\usepackage{graphicx}
\usepackage{subcaption}
\usepackage{dcolumn}
\usepackage{bm}
\usepackage{natbib}
\usepackage{color}
\usepackage{booktabs}
\usepackage{relsize}

\newcommand{\ra}{\rightarrow}

\newcommand{\lap}{\mbox{$\cal L$}}

\newcommand{\pr}{\mbox{Pr}}

\renewcommand{\thetable}{\arabic{table}}
\newcommand{\beginsupplement}{%
        \setcounter{table}{0}
        \renewcommand{\thetable}{S\arabic{table}}%
        \setcounter{figure}{0}
        \renewcommand{\thefigure}{S\arabic{figure}}%
     }
\newtheorem{theorem}{Theorem}
\begin{document}

\preprint{APS/123-QED}

\title{A Jarzynski-type equality for nonequilibrium steady-state probabilities}

\author{U\u{g}ur \c{C}etiner} 
 \email{cetiner.ugur@gmail.com}
\affiliation{%
Department of Systems Biology, Harvard Medical School, Boston, Massachusetts 02115, USA
}%
\begin{abstract}
We show that steady-state probabilities of a nonequilibrium Markovian system can be reconstructed from a weighted ensemble average of finite-time loop-erased paths. Each path $\Gamma$ is weighted by $e^{-S(\Gamma)}$, where $S(\Gamma)$ can be interpreted either as the action functional or as the entropy change in the surrounding reservoirs. In doing so, we uncover a Jarzynski-type equality for nonequilibrium steady states. We also reveal that the probabilities of loop-erased paths obey strong thermodynamic symmetry relations. 
\end{abstract}
\maketitle
\section*{Introduction} 
Consider a thermodynamic process that connects two equilibrium states of a system by doing work ($W$) on it. For example, grab a rubber band whose initial length is $A$ units. Assume that the rubber band has been at rest in a room with a temperature $T$. So, initially, it is in an equilibrium state characterized by $(A,T)$. Next, stretch the rubber band to a new length $B$ and allow it to relax into a new equilibrium state characterized by ($B,T$). The system has been driven away from an initial equilibrium state by changing a control parameter according to a protocol. Here, the control parameter is the length of the rubber band and we do work by changing it, typically away from thermal equilibrium. The second law of thermodynamics tells us that 
\begin{equation}\label{secondlaw}
W\geq \Delta F,
\end{equation}
where $\Delta F$ is the free energy difference between two equilibrium states, $\Delta F=F_{B,T}-F_{A,T}$. Eq.\ref{secondlaw} becomes an equality when the process is executed extremely slowly, ensuring that the rubber band consistently remains in thermal equilibrium with the room throughout the stretching. Such a \emph{quasi-static} process recovers $\Delta F$ exactly but necessarily takes infinite time.
Let us repeat the same stretching exercise again, but with a nanoscale rubber band. 
Due to molecular-level jiggling in the reservoir, the amount of work done on the nanoscopic rubber band changes from one realization to the next even though we use the same stretching protocol. If we repeat this process many times, we would report a distribution of work values, and the second law now becomes a statistical statement,
\begin{equation}
\langle W \rangle \geq \Delta F,
\end{equation}
where the angle brackets represent the average over the work distribution. Such nanoscale experiments are a daily practice in the study of single molecules, and they are typically conducted with the help of optical tweezers or patch-clamp apparatus \cite{liphardt2002equilibrium,collin2005verification,douarche2005experimental,ritort2006single,ccetiner2020recovery,
ccetiner2023dissipation,ccetiner2024mechanosensitive}. For the rest of the paper, we set $k_B=T=1$, and work with dimensionless entropy and energy. Here, $k_B$ is the Boltzmann constant and $T$ is the ambient temperature.

 In 1997, Jarzynski showed that nanoscale fluctuations in driven systems encode important equilibrium information, which can be recovered in finite time \cite{jarzynski1997nonequilibrium}. In particular, the free energy difference between two equilibrium states ($\Delta F$) can be exactly recovered from an ensemble of driven nonequilibrium trajectories between the two states. To do this, each trajectory should be \emph{weighted} by $\mathrm{e}^{-W}$, where $W$ is the work performed on the system during the trajectory \citep{jarzynski2001does}. Specifically,  
\begin{equation}\label{Jarzynski}
\mathrm{e}^{-\Delta F}=\lim_{N\rightarrow \infty} \frac{1}{N} \sum_{k=1}^N \mathrm{e}^{-W_k}\equiv \langle \mathrm{e}^{-W} \rangle \,,
\end{equation}
where $W_k$ is the work performed on the system during the $k$-th realization of the ensemble and $N$ is the size of the ensemble. Note that the left-hand side of the equality above contains the equilibrium information ($\Delta F$) whereas the right-hand side in general represents a nonequilibrium ensemble. That is, unless the control parameter undergoes a quasi-static change over an infinite time, the system is driven away from equilibrium during each finite-time trajectory.

Here, we derive an analogue of the Jarzynski equality in Eq.\ref{Jarzynski} for nonequilibrium steady-state probabilities. We discuss the key findings here, reserving detailed proofs for the Results and Supplementary Information. We consider a continuous-time, time-homogeneous Markov process, $X(t)$, on a discrete state space $V=\{1,\dots,n\}$. In the \emph{linear framework}, the Markov process is represented by a directed graph with labeled edges, denoted as $G$ \citep{gunawardena2012linear,mirzaev2013laplacian}; for recent reviews, see \cite{nam2022linear,kjg23}. Graph vertices represent the states of the Markov process, directed edges, denoted $i \to j$, represent transitions with positive infinitesimal rates. The edge labels show these rates. For thermodynamic reasons, we will assume that graphs are reversible, so that if $i \to j$ then also $j \to i$. We also assume local detailed balance and that $G$ is strongly connected, so that any two distinct vertices are connected by a directed path, which implies that the Markovian dynamics relaxes to a unique steady state \cite{nam2022linear}. In the linear framework, this steady state may be calculated, up to a scalar multiple, as a vector $\rho(G)$, where each $\rho_i(G)$ is a rational algebraic function of the edge labels \cite{nam2022linear}. If $p_i$ is the steady-state probability of being in state $i$, then $p_i=\lambda \rho_i(G)$, where $\lambda$ is the proportionality constant, which can be removed by normalizing to the total probability. Hence, knowing $\rho(G)$ is equivalent to knowing the steady-state probability $p$.

The probability $p_i$ can be obtained by measuring the proportion of time that a typical trajectory spends in state $i$, in the limit of an infinite trajectory. By analogy with the Jarzynski-equality in Eq.\ref{Jarzynski}, we find that $p_i$ can also be obtained as an average over an infinite ensemble of finite-time trajectories, in which loops have been removed. Choose an arbitrary reference vertex in the underlying graph $G$, usually taken to be $1$. Create an ensemble of stochastic paths (trajectories) starting from vertex $i$ and terminating when they first reach $1$. (Some paths may never reach $1$ but these form a set of measure zero and may be ignored.) Delete the loops in each path by using the loop-erasure algorithm described later. Let $\Gamma_k$ denote the loop-erased path formed during the $k$-th realization of the ensemble. Weight each loop-erased path by $\mathrm{e}^{-S(\Gamma_k)}$, where $S(\Gamma_k)$ is the total change in the \textit{entropy} of the surrounding reservoirs along the loop-erased path (Eq.\ref{e-spp}). $S(\Gamma_k)$ is also known as the \emph{action functional} \cite{lebowitz1999gallavotti}. The steady-state probabilities can be recovered from the following equality, 
\begin{equation}\label{Result1}
\rho_i(G)= \lim_{N\rightarrow \infty} \frac{1}{N} \sum_{k=1}^N \mathrm{e}^{-S(\Gamma_k)}\equiv \langle \mathrm{e}^{-S} \rangle.
\end{equation} 
We will use Wilson's algorithm \cite{wilson1996generating} and our previous reformulation of nonequilibrium steady-state probabilities \cite{ccetiner2022reformulating} to prove Eq.\ref{Result1}.


It is interesting that both Eq.\ref{Jarzynski} and Eq.\ref{Result1} share the same mathematical structure, but in the latter, $W$ is replaced by $S$. While a time-homogeneous Markov process is not a driven system, the way these formulas recover useful information suggests the following analogy. In driven systems, the quantity of interest is the free energy difference between two equilibrium states, $\Delta F$, whereas for a time-homogeneous Markov process, it is the nonequilibrium steady-state probability, $p_i$. In both cases, we know how to obtain these quantities from a single trajectory in the \emph{infinite-time} limit. As noted above, if we change the control parameter quasi-statically, the limiting work done is equal to $\Delta F$, while if we measure the limiting proportion of time spent in state $i$, we recover $p_i$. The question arises of whether we can recover the same information with an ensemble of \emph{finite-time} trajectories. We can, provided the trajectories are suitably weighted. In the case of driven systems, the weight is $\mathrm{e}^{-W}$, while for Markov processes, the weight is $\mathrm{e}^{-S}$. Eq.\ref{Jarzynski} remains independent of the rate at which the control parameter is changed, as long as it connects the same points in the parameter space. Similarly, Eq.\ref{Result1} is unaffected by the time it takes for a stochastic trajectory to reach the reference vertex, provided it does reach it.

In the quasi-static limit, in which the driven system traces a set of equilibrium states throughout the protocol, $W$ becomes a path-independent quantity. Similarly, $S$ becomes path-independent at thermodynamic equilibrium (see below). Away from equilibrium, these variables fluctuate from one realization to the next. Nevertheless, both Eq.\ref{Jarzynski} and Eq.\ref{Result1} recover the needed information through the appropriate weighting. In doing so, they trade one infinity---time---with another infinity---the size of the ensemble. 

The idea of loop erasure that we exploit here also implies an unexpected thermodynamic symmetry. Consider an ensemble of stochastic trajectories starting at vertex $i$ and terminating when they first hit vertex $j$. Once we erase loops from a trajectory, we are left with a directed path from vertex $i$ to $j$ in which all vertices are distinct. We call such paths \emph{minimal} and denote the set of minimal paths from $i$ to $j$ by $M(i,j)$ \cite{ccetiner2022reformulating}. For a finite graph, the set $M(i,j)$ is always finite. The ensemble of loop-erased trajectories yields a probability distribution on $M(i,j)$ as follows. If $\Gamma \in M(i,j)$, then $\Pr(\Gamma)$ is the limiting proportion of trajectories in the ensemble that yield $\Gamma$ after loop erasure. Reversing the direction of edges on $\Gamma \in M(i,j)$ yields another minimal path, $\hat{\Gamma}\in M(j,i)$ and defines a bijection between these sets. Note that $M(j,i)$ also acquires a probability distribution from loop erasure. If the Markov process reaches a steady state of thermodynamic equilibrium, then we show that,
\begin{equation}\label{Result3}
\Pr(\Gamma)=\Pr(\hat{\Gamma}),
\end{equation}
which is a form of time-reversal symmetry coming from loop erasure. However, if the system reaches a nonequilibrium steady state, this time-reversal symmetry is broken and Eq.\ref{Result3} no longer holds. Instead, we have the following relation governing the probability distributions on minimal paths,
\begin{equation}\label{Result3b}
    \frac{\Pr(\Gamma)}{\Pr(\hat{\Gamma})}=\mathrm{e}^{S(\Gamma)}\left(\frac{p_i}{p_j} \right).
\end{equation}
If the system reaches thermodynamic equilibrium, the steady-state probabilities follow the Boltzmann distribution, so that $p_i/p_j=\mathrm{e}^{-(F_i-F_j)}=\mathrm{e}^{-\Delta F}$ and, furthermore, the path entropy $S(\Gamma)$ has the same value, so that $S(\Gamma) = \Delta F$. Hence, Eq.\ref{Result3b} reduces to Eq.\ref{Result3} for systems that reach thermodynamic equilibrium.
 
Eqs.\ref{Result3} and \ref{Result3b} establish previously unsuspected connections between probabilities on combinatorial structures of linear framework graphs and thermodynamics. We provide proofs of these results in the Supplementary Information.
\section*{SETUP}
We investigate the dynamics of a system governed by a continuous-time, finite-state, time-homogeneous Markov process, denoted as $X(t)$. The state space is given by $V=\{1, 2, \dotsc, n\}$. We use $l(i \to j)$'s to represent the infinitesimal transition rates with dimensions of (time)$^{-1}$,
\begin{equation}\label{Ratedef}
    \ell(i \to j) = \lim_{\Delta t \to 0}\frac{\Pr(X(t+\Delta t) = j \mid X(t) = i)}{\Delta t} \,,
\end{equation}
where the numerator is the probability of being in state $j$ at time $t+\Delta t$ given that the system was in state $i$ at time $t$. As mentioned before, Markov processes can be represented by graphs. The linear framework exploits this correspondence between graphs and Markov processes and has demonstrated its utility in the investigation of nonequilibrium systems \citep{gunawardena2012linear,cetineruniversal}. Letting $p_i(t)$ denote the shorthand notation for $\Pr(X(t) = i)$, the evolution of the system is described by the master equation \citep{mirzaev2013laplacian,van1992stochastic},
\begin{equation}\label{Master}
    \frac{dp(t)}{dt} = {\cal L}(G) \cdot p(t) \,,
\end{equation}
where $p(t) = (p_1(t), \dotsc, p_n(t))^T$ is an $n\times 1$ column vector and ${\cal L}(G)$ is the $n\times n$ Laplacian matrix of the graph. The entries of the Laplacian are given by, 
\begin{equation}\label{Lmatrix}
    {\cal L}(G)_{ji} = \begin{cases}
        \ell(i \to j) & \text{if $i \neq j$} \\
        -\sum_{k \neq i}{\ell(i \to k)} & \text{if $i=j$.}
    \end{cases}
\end{equation}

Consider a stochastic trajectory denoted as $\{X(t)\}_{t\geq 0}$ and represent the sequence of states visited in this trajectory as a path on $G$, which we denote as $P = [i_1, i_2, \dots, i_{k-1}, i_k]$. A path is said to be \emph{minimal} if all of its vertices (states) are distinct---It is important to keep in mind that minimal paths can have different names in different fields. They are sometimes called simple paths or self-avoiding paths as well---Paths can be represented as directed paths by indicating directed edges between adjacent vertices. For example, the directed path representation of $P$ is $P: i_1 \to i_2 \to \dots \to i_{k-1} \to i_k$. Here, $i_1$ is referred to as the initial vertex, and $i_k$ is the terminal vertex. We represent the reversed path with $\hat{P}$, $\hat{P}:i_k\ra i_{k-1}\dots \ra i_2\ra i_1$. If the initial and terminal vertices of a path are the same, the path forms a cycle. The set of all minimal paths from vertex $i$ to vertex $j$ is represented by $M(i,j)$. 

Another graph-theoretical concept is a \emph{tree}, which is a subgraph of $G$ that contains no cycles of edges. When a tree contains all the vertices of $G$, it is called a \emph{spanning tree}. Moreover, a spanning tree is considered \emph{rooted} at state $i$ if it lacks any edges departing from vertex $i$. See Fig.\ref{Fig1} for some examples of rooted spanning trees. We can assign a positive real number, or ``weight", to each spanning tree by utilizing the function $q(H)$. Here, $q(H)$ represents the product of the edge labels on the edges of subgraph $H$, calculated as $q(H) = \prod_{i \rightarrow j \in H} \ell(i \rightarrow j)$. We also extend $q$ to the subset of spanning trees by defining $q(X)=\sum_{T\in X} q(T)$, where $X$ is a set of spanning trees.

While the master equation captures the system's dynamics, it lacks a thermodynamic interpretation in the absence of additional assumptions. To this end, we assume that the log ratio, $\ln[\ell(i \to j) / \ell(j \to i)]$ is the total entropy change induced by the transition from $i$ to $j$, i.e., the entropy change in the reservoirs together with the internal entropy difference between $j$ and $i$. This assumption is called local detailed balance, and it was first introduced in 1983 by Katz, Lebowitz, and Spohn in their examination of nonequilibrium steady states in a lattice gas composed of interacting particles \citep{katz1983phase}. This concept, however, has its historical roots traced back to the work of Hill and Schnakenberg \citep{hill1966studies,schnakenberg1976network}. Since then, it has played a fundamental role in stochastic thermodynamics \citep{seifert2005entropy, bauer2014local}. For a directed path $P: i_1 \to i_2 \to \dots \to i_{k-1} \to i_k$, the total entropy change along the path is given as,
\begin{equation}
S(P) = \ln\left[\left(\frac{\ell (i_1 \to i_2)}{\ell(i_2 \to i_1)}\right) \cdots \left(\frac{\ell(i_{k-1} \to i_k)}{\ell(i_k \to i_{k-1})}\right)\right] \,.
\label{e-spp}
\end{equation}
Eq.\ref{e-spp} may also be called action functional \cite{lebowitz1999gallavotti}. If the path $P$ is a cycle, then $S(P)$ is called the thermodynamic affinity of $P$ \cite{schnakenberg1976network}. 
\section*{PRELIMINARY RESULTS}
\subsection*{Thermodynamic equilibrium}
The system reaches steady state when the probabilities of occupying each state no longer change over time. Eq.\ref{Master} reveals that any solution $p$ that satisfies this steady-state condition must reside within the kernel of the matrix $p \in \ker{{\cal L}(G)}$. For a strongly-connected graph, the kernel is one-dimensional. Consequently, the uniqueness of $p$ is guaranteed, as it must also satisfy the normalization condition $\sum_{i=1}^n{p_i} = 1$. 

We say that the system is in thermodynamic equilibrium if each transition is balanced with the reversed one, $p_i \, \ell(i \to j) = p_j \, \ell(j \to i)$ for all $i,j \in V$. In this case, the system is described by the equilibrium Boltzmann distribution. We also say that time-reversal symmetry holds. Equivalently, $S(C) = 0$ holds for every cycle $C$. This condition, referred to as the \emph{cycle condition}, is both necessary and sufficient for achieving thermodynamic equilibrium. Importantly, if the cycle condition holds, $S$ does not depend on the specific path taken between states $i$ and $j$. With these insights, we can now proceed to calculate a basis element $\mu(G) \in \ker{{\cal L}(G)}$ at thermodynamic equilibrium. Assuming the reference vertex to be vertex $1$, we can select any path $P_i \in M(i,1)$ and assign $\mu_i(G) = \mathrm{e}^{-S(P_i)}$. By construction, $\mu_1(G)=1$. Given that $\mu(G) \in \ker{{\cal L}(G)}$ and the $\ker{{\cal L}(G)}$ is one-dimensional, it follows that 
\begin{equation}
p_i \propto \mu_i(G) = \mathrm{e}^{-S(P_i)},
\label{e-mu}
\end{equation}
for any $P_i \in M(i,1)$. The proportionality constant can determined from the normalization condition:
\begin{equation}\label{normalized}
p_i = \frac{\mathrm{e}^{-S(P_i)}}{\sum_j \mathrm{e}^{-S(P_j)}}.
\end{equation} 
\subsection*{The arboreal distribution and reformulation}
If the system is not at thermodynamic equilibrium, $S$ becomes path-dependent, so a unique solution cannot be constructed using Eq.\ref{normalized}. In such situations, we can derive alternative basis elements by using a probability distribution defined on spanning trees rooted at the reference vertex. This distribution, which we refer to as the \emph{arboreal distribution}, is connected to the probability measures generated by loop-erased random walks on spanning trees, as shown in the Results.

Let $\Theta_i$ denote the set of spanning trees of $G$ that are rooted at vertex $i$. We introduce a probability distribution on the spanning trees in $\Theta_1$ as follows, 
\begin{equation}\label{arboreal}
\pr_{\Theta_1}(T) = \frac{q(T)}{\sum_{T \in \Theta_1} q(T)}=\frac{q(T)}{q(\Theta_1)}.
\end{equation}
That is, the probability assigned to a spanning tree $T \in \Theta_1$ is directly proportional to its weight $q(T)$ \citep{ccetiner2022reformulating}. This is the arboreal distribution. By construction, there is a unique minimal path from $i$ to the root on every rooted spanning tree, which is denoted by $T_i$. Let $S(T_i)$ be the corresponding path entropy or action, then an alternative basis element, $\rho(G) \in \ker{\cal L}(G)$, is given by,
\begin{equation}
 \rho_i(G)=\sum_{T \in \Theta_1} \pr_{\Theta_1}(T)\mathrm{e}^{-S(T_i)}\equiv\langle \mathrm{e}^{-S(T_i)}\rangle_{\Theta_1},
 \label{rhoness}
\end{equation}
where the angle brackets denote an average taken over the arboreal distribution. See the Supplement for a derivation. Similar to Eq.\ref{e-mu}, the steady-state probabilities are recovered using the normalization condition: 
\begin{equation}\label{NessFinal}
p_i=\frac{\langle \mathrm{e}^{-S(T_i)}\rangle_{\Theta_1}}{\sum_j \langle \mathrm{e}^{-S(T_j)}\rangle_{\Theta_1}}.
\end{equation}
It is worth noting that if the system is at thermodynamic equilibrium, Eqs.\ref{rhoness} and \ref{NessFinal} can be simplified to Eqs.\ref{e-mu} and \ref{normalized}, respectively. In doing so, we recover both the notion of path independence and the equilibrium formalism.
\subsection*{Loop-Erased Random Walk}
 Lawler introduced the loop-erased random walk in 1980 \citep{lawler1980self}, and it has since been used in a variety of disciplines, including combinatorics, computer science, and statistical physics. LERW exhibits intriguing properties, including conformal invariance and its connection to critical percolation, making it a valuable tool for studying scaling limits and investigating the behavior of finite geometries \citep{pemantle1991choosing,cardy1992critical,wilson1996generating,
propp1998get,schramm2000scaling,lubeck2004universal,lawler2011conformal,
lawler2013intersections,wiese2019field}. Here is an algorithm for loop erasure \citep{lawler1980self}. Let $P=[i_1,i_2,\dots,i_k]$ be a path on $G$. The corresponding loop-erased path $\Gamma$=LE($P$) is obtained from $P$ by erasing cycles on $P$ as follows. First, set $u_1\equiv i_1$, and $l_1\equiv \text{max}\{j: 1\leq j\leq k, i_j=u_1\}$. If $l_1=k$, $\Gamma$=$[u_1]$ and the process terminates. As long as $l_i<k$, set $u_{i+1}=i_{l_{i}+1}$ and obtain $l_i$'s inductively $l_{i+1}\equiv \text{max}\{j: l_i\leq j\leq k, i_j=u_{i+1}\}$. Stop when $l_m=k$. The loop-erased path is a minimal path with the same initial and terminal vertices of $P$, $\Gamma=\text{LE}(P)=[u_1=i_1,\dots,u_m=i_k]$. $\Gamma$ can also be represented as a directed path $u_1\to\dots \to u_m$. For example, if a path on $G$ is $P:2\to3\to2\to1$, the loop erasure algorithm returns LE($P$)$: 2\to1$. LERW is also used in the Wilson algorithm, which outputs a random spanning tree of a graph, as we will discuss next
\subsection*{Wilson's Algorithm}
In 1996, Wilson utilized loop-erased random walks to generate spanning trees of a given graph uniformly at random without explicitly enumerating them \citep{wilson1996generating}. Below, we adopt a slightly different version of the Wilson algorithm. The following algorithm outputs a random spanning tree rooted at the reference vertex, according to the arboreal distribution $\Pr_{\Theta_1}(T)$. 
\begin{enumerate}
\item Given that vertex 1 is the reference vertex, let $A_1=\{1\}$ and $V_1= V\setminus A_1= \{2,3,\dots,n\}$. Initiate a stochastic trajectory from the smallest vertex in $V_1$ and stop when the trajectory hits $A_1$. Delete all the loops in this path. If $P = [i_1=2, i_2, \dots, i_k=1]$ and $\Gamma=\text{LE}(P)=[u_1,\dots,u_m]$, include all the edges that are visited by LERW in the tree, e.g., $u_1\to\dots\to u_m=1$, constructing the backbone of the tree. Set $A_2=A_1\cup\{u_1,\dots,u_m\}$ and let $V_2=V\setminus A_2$ be the set of vertices that have not been added to the tree. Do the following recursively,
\item If $V_k=\emptyset$, then we have a spanning tree rooted at 1, and we stop. Otherwise, choose the smallest vertex in $V_k$ and take a LERW from that vertex to $A_k$. Add the edges obtained from LERW to the existing tree and let $V_{k+1}=V\setminus A_k$ be the set of vertices that are not on the tree.
\end{enumerate} 
This algorithm halts with probability one \citep{wilson1996generating}. The Wilson algorithm has been proven using different techniques. Wilson utilized the concept of \emph{cycle popping} to establish the correctness of his algorithm \citep{wilson1996generating}, while Lawler employed the Green functions in his proof \citep{lawler2013intersections,loopmeasures2018}. The Supplement includes a brief account of Lawler's method, which establishes the connection between the Wilson algorithm and the arboreal distribution. While these arguments are not necessary to follow our results, we have provided them for completeness.
\vspace{-1cm}
\section*{Proof of the main result}
To reconstruct the steady-state solution, we first create an ensemble of LERWs from $i$ to 1, and enumerate each realization as $\Gamma_k$ for $k=1,2,\dots N$, where $N$ be the total number of such trajectories. This ensemble induces a probability distribution on $M(i,1)$ such that the probability of observing $m \in M(i,1)$ is given by
\begin{equation}\label{P(m)ensemble}
\text{Pr}(m)=\lim_{N \to \infty}\frac{1}{N}\sum_{k=1}^N \delta(\Gamma_k-m)=\langle \delta(\Gamma_k-m) \rangle,
\end{equation}
where $\delta$ is the Kronecker delta function: $\delta(\Gamma_k-m)=1$ if $\Gamma_k=m$ and zero otherwise. The angle brackets represent an ensemble average, as commonly used. Given that there exists a unique minimal path from vertex $i$ to 1 on every spanning tree rooted at 1, we define $T(m)=\{T\in \Theta_1| T_i=m\}$ as the set of spanning trees rooted at 1, where the unique minimal path $T_i$ is equal to $m$. With this in mind, using the Wilson algorithm, Eq.\ref{P(m)ensemble} can be related to the arboreal distribution as follows,
\begin{equation}\label{P(m)}
\text{Pr}(m)= \sum_{T \in T(m)} \pr_{\Theta_1}(T).
\end{equation}
That is, the probability of generating a certain minimal $m$ in an ensemble of LERW's from  $i$ to 1 is equal to the total arboreal probability of spanning trees that have $m$ as the unique minimal path from $i$ to 1. To show this, consider generating random spanning trees rooted at the reference vertex by using the Wilson algorithm, which gives the joint probability of having $m$ and $T$ as,
\begin{equation}
 \pr(m,T)= 
\begin{cases}
    \pr_{\Theta_1}(T),& \text{if}\ T_i=m\\
    0,              & \text{otherwise}.
\end{cases}
\end{equation}
Accordingly, $\pr(m)=\sum_{T\in \Theta_1} \pr(m,T)$, which results in Eq.\ref{P(m)} (also, see Corollary 3.6 in \citep{hu2021reverse}).  In the Supplement, we show Eq.\ref{P(m)} in action for the three-state graph depicted in Fig.\ref{Fig1}. Eq.\ref{P(m)} offers an interpretation of the arboreal distribution in terms of LERW probabilities. Using this connection, we can prove the main result, Eq.\ref{Result1}, as follows,
\begin{equation} 
\begin{split}
 \rho_i=&\sum_{T \in \Theta_1} \pr_{\Theta_1}(T)e^{-S(T_i)}=\sum_{m \in M(i,1)} \pr(m)e^{-S(m)}\\ 
 &= \sum_{m \in M(i,1)} \lim_{N \to \infty}\frac{1}{N}\sum_{k=1}^N \delta(\Gamma_k-m)e^{-S(m)}\\
 &=\lim_{N \to \infty}\frac{1}{N}\sum_{k=1}^N \sum_{m \in M(i,1)} \delta(\Gamma_k-m) e^{-S(m)}\\
 &=\lim_{N \to \infty}\frac{1}{N}\sum_{k=1}^Ne^{-S(\Gamma_k)}= \langle e^{-S(\Gamma)} \rangle.
 \end{split}
\end{equation}
In the first line, the steady-state solution is represented as a summation over minimal paths $m\in M(i,1)$. In the second line, we substituted the ensemble definition of $\pr(m)$, as given by Eq.\ref{P(m)ensemble}. Transitioning to the third line, we exchanged the order of the summations, and utilized the definition of the Kronecker delta function, leading us to the final line, which is Eq.\ref{Result1}.

\vspace{-0.4cm}
\section*{Conclusions}
In driven systems, the equilibrium free energy difference can be calculated from a single quasi-static process. For time-homogeneous Markovian systems, the steady-state probability $p_i$ can be obtained by measuring the proportion of time that a typical trajectory spends in state $i$. Both require infinite time. However, if we use the correct statistical weights, we can recover these quantities from finite-time ensembles. The idea of using weighted averages to recover useful information from finite-time trajectories is a common theme in nonequilibrium statistical mechanics. Hummer and Szabo established a connection between the weighting procedure and the Feynman-Kac theorem of stochastic differential equations \citep{hummer2001free}. Many well-known fluctuation theorems can also be derived as a special case of path-ensemble averages, as demonstrated by Crooks \citep{crooks2000path}. It is also interesting to see that the probabilities defined on loop-erased paths inherit a strong symmetry relation from the time-reversal symmetry of the underlying Markov process. Even in the absence of time-reversal symmetry, thermodynamics remains a decisive factor in shaping these probabilities. An important research direction is to explore the extent of these analogies and establish further connections with other equalities in nonequilibrium steady states.
\section*{Acknowledgment}
UC thanks Chris Jarzynski and Oren Raz for discussions. UC is indebted to Jeremy Gunawardena for the ongoing support and guidance. 
\bibliography{Ness_LERW}
\newpage
\onecolumngrid
\appendix
\beginsupplement
\section*{Supplementary Information}
\subsection*{The Linear Framework}
\begin{figure}
\includegraphics[scale=0.5]{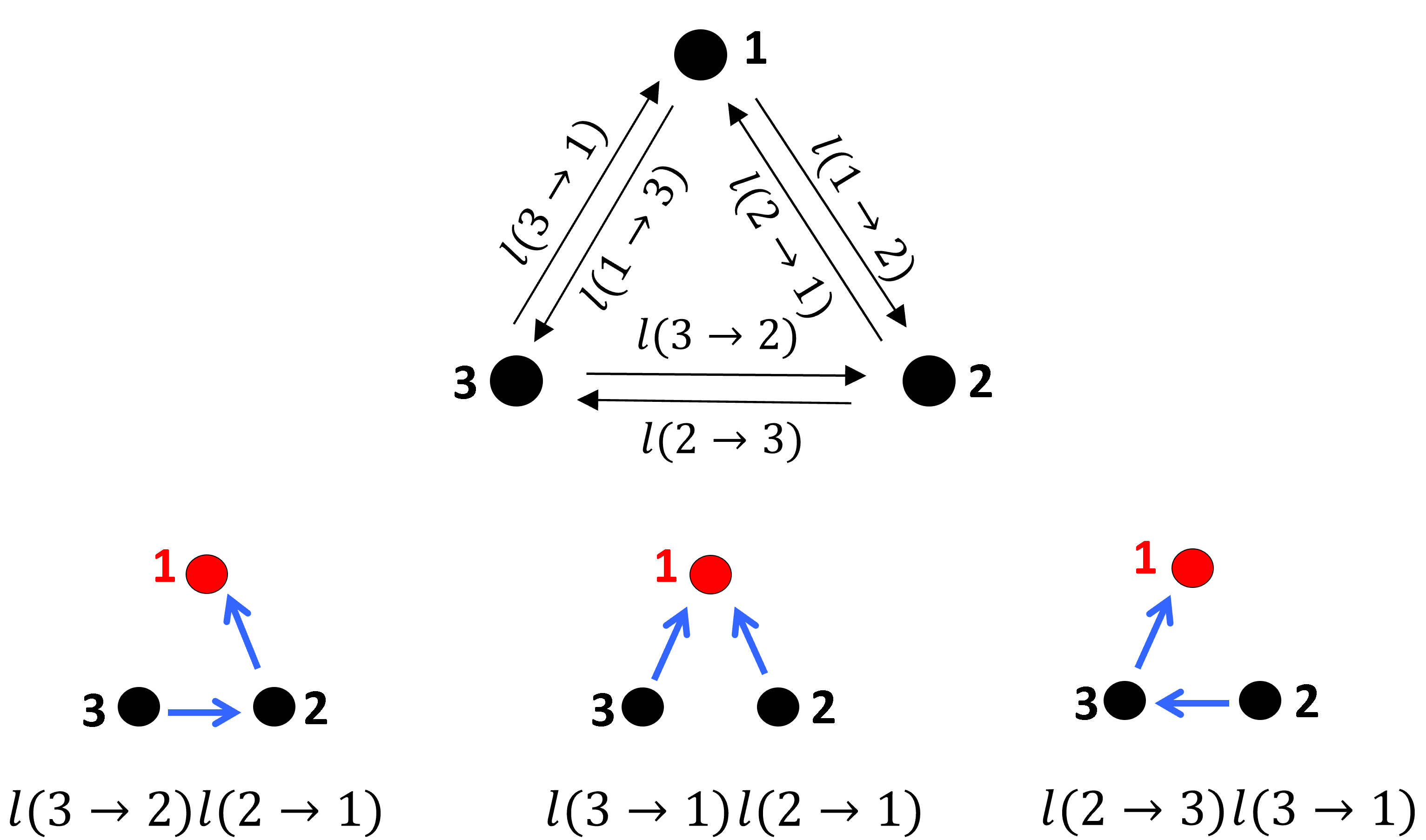}
\caption{The graph representation of a three-state Markov process, where each vertex represents a state, edges represent possible transitions, and edge labels indicate transition rates. In the illustration below, the spanning trees rooted at vertex 1 are shown, with the root vertex highlighted in red. Spanning trees can be assigned weights, calculated as the product of the edge labels on each tree. The corresponding weights for each tree are provided below the illustration.}
\label{Fig1}
\end{figure}

We provide a brief review of the linear framework, which serves as a foundation for Eq.\ref{rhoness}. More details and background can be found in \cite{gunawardena2012linear,mirzaev2013laplacian}. The linear framework is a graph-theoretical method that expresses the steady-state solution of Eq.\ref{Master} as rational algebraic functions of the transition rates. The linear framework derives its name from the linearity of the master equation. An alternative basis element in the linear framework, $\tilde{\rho}\in \ker {\cal L}(G)$, is given by the following expression,
\begin{equation}\label{lf}
\tilde{\rho}_i(G)=\sum_{T\in \Theta_i} q(T).
\end{equation}
Here, $\Theta_i$ is the set of spanning trees rooted at $i$ and $q(T)=\prod_{i \ra j \in T} \ell(i\ra j)$. A strongly connected graph has a spanning tree rooted at each vertex, so that $\Theta_i \neq \emptyset$. For example, Fig.\ref{Fig1} displays $\Theta_1$ for the three-state graph given in the same figure. The steady-state solution can be obtained using the normalization condition,
\begin{equation}
p_i = \frac{\tilde{\rho}_i(G)}{\tilde{\rho}_1(G) + \cdots + \tilde{\rho}_n(G)} \,,
\label{lf2}
\end{equation}
where $n$ is the number of states. There are two issues that limit the applicability of this elegant expression. First, the number of spanning trees grows \emph{super-exponentially} with the size of the graph and all of these trees are required to calculate steady-state probabilities exactly. This gives rise to the algebraic explosion in steady-state calculations. Also, the rational function lacks any thermodynamic interpretation. Eq.\ref{rhoness}, has been formulated to address these issues \citep{ccetiner2022reformulating} and used as canonical basis element for $\ker \lap(G)$ in the main text. A derivation can be given by first rewriting Eq.\ref{lf} in terms of path entropies or action \cite{ccetiner2022reformulating},
\begin{equation}
\tilde{\rho}_i(G)=\sum_{T \in \Theta_1} \mathrm{e}^{-S(T_i)}q(T).
\end{equation}
From this equation, it follows that $\tilde{\rho}_i(G)\propto \langle \mathrm{e}^{-S(T_i)} \rangle_{\Theta_1}$, where the average is
taken over the arboreal distribution. Since the steady state is proportional to $\tilde{\rho}_i$, we obtain the following relation,
\begin{equation}
p_i \propto  \langle \mathrm{e}^{-S(T_i)} \rangle_{\Theta_1},
\end{equation}
which gives another choice for the basis element for the kernel of \lap(G).
\subsection*{Proof of Eq.\ref{Result3}}
We first claim that the arboreal distribution does not depend on the choice of the reference vertex at thermodynamic equilibrium. To show this, we use a bijective function $\Phi_{i,j}\ : \Theta_i\rightarrow \Theta_j$, defined as follows \cite{wong2018structural}. Pick a spanning tree rooted at state $i$, $T \in \Theta_i$ and reverse the edges on the unique minimal path from $j$ to $i$ on this tree. This process yields a spanning tree rooted at $j$, which is $\Phi_{i,j}(T)\in \Theta_j$. In Fig.\ref{Fig-bijection}, as an example, we show how this bijection works for the spanning trees of the three-state graph in Fig.\ref{Fig1}. Note that $q\left(\Phi_{i,j}(T)\right)=\mathrm{e}^{-S(T_j)}q(T)$. Using this equality, and the fact that at thermodynamic equilibrium, path entropies (or actions) do not depend on the path taken between any two vertices, we can show that the probability of a spanning tree $T \in \Theta_i$ is the same as the probability of the corresponding spanning tree $\Phi_{i,j}(T)\in \Theta_j$. In the first case, the arboreal distribution is defined on $\Theta_i$ and in the second case, it is defined on $\Theta_j$.  Starting from the definition of the arboreal distribution for spanning trees rooted at $j$, we obtain the following result,
\begin{equation} \label{arborealsymmetry}
\begin{split}
\pr_{\Theta_j}(\Phi_{i,j}(T)) &= \frac{q(\Phi_{i,j}(T))}{\sum_{\Phi_{i,j}(T) \in \Theta_j} q(\Phi_{i,j}(T))}\\
&= \frac{q(T)}{\sum_{T \in \Theta_i} q(T)}=\pr_{\Theta_i}(T).
\end{split}
\end{equation}
In the second line, as mentioned earlier, $S(T_j)$ does not depend on the path taken between $i$ and $j$, therefore, $\mathrm{e}^{-S(T_j)}$ is a common factor both in the numerator and all terms in the denominator, so it cancels out. The fact that the arboreal distribution remains unaffected by the choice of the reference vertex reveals how the time-reversal symmetry of the underlying Markovian dynamics extends into the probabilities defined on the space of rooted spanning trees. This serves as a pivotal point for exploring how this time-reversal symmetry translates to the level of probabilities defined on loop-erased paths, as we show next.

Let $\Gamma \in M(i,j)$ and $T(\Gamma)=\{T\in \Theta_j\ |\ T_i=\Gamma\}$. Consider $\hat{\Gamma} \in M(j,i)$ and let $T(\hat{\Gamma})=\{\Phi_{j,i}(T)\ |\ T \in T(\Gamma)\}$. 
Create two ensembles of loop-erased trajectories, one from $j$ to $i$ and another one from $i$ to $j$. The probability $\Gamma$ is given by Eq.\ref{P(m)},
\begin{equation}\label{SIproof1}
\begin{split}
\pr(\Gamma)=&\sum_{T \in T(\Gamma)} \pr_{\Theta_j}(T)\\
=& \sum_{T \in T(\Gamma)}  \pr_{\Theta_i}(\Phi_{j,i}(T))=\pr(\hat{\Gamma}).
\end{split}
\end{equation}
Moving from the first line to the second, we utilized the bijection and exploited the equilibrium symmetry of the arboreal distribution. This completes the proof of Eq.\ref{Result3}. Another proof can be given using certain properties of the underlying Green function of the Markov process \cite{loopmeasures2018}. Here, we provide a new proof by using graph-theoretical tools. 
\subsection*{Proof of Eq.\ref{Result3b}}
If the system operates at a nonequilibrium steady state, these symmetries originating from time-reversal symmetry break down. But, even away from equilibrium, there are interesting relations between these path probabilities, as we demonstrate here. For $\Gamma \in M(i,j)$, consider its probability under the ensemble of loop-erased trajectories from $i$ to $j$,
\begin{equation}\label{Prob1}
    \pr(\Gamma)=\sum_{T \in T(\Gamma)} \pr_{\Theta_j}(T)=\frac{q(T(\Gamma))}{q(\Theta_j)},
\end{equation}
where $T(\Gamma)=\{T\in \Theta_j\ |\ T_i=\Gamma\}$ as defined before. By construction, the weights of the trees in $T(\Gamma)$ have a common factor of $q(\Gamma)$, therefore, $q(T(\Gamma))=q(\Gamma)\alpha$, where $\alpha$ is some constant, representing the rest of the contribution to $q(T(\Gamma))$ from the other branches on these trees. Similarly, 
\begin{equation}\label{Prob2}
    \pr(\hat{\Gamma})=\sum_{T \in T(\Gamma)} \pr_{\Theta_i}(\Phi_{j,i}(T))=\frac{q(T(\hat{\Gamma)})}{q(\Theta_i)},
\end{equation}
where $T(\hat{\Gamma})=\{\Phi_{j,i}(T)\ |\ T \in T(\Gamma)\}$. The weights of the trees in  $T(\hat{\Gamma})$ have
a common factor of $q(\hat{\Gamma})$, moreover, it is not hard to see that $q(T(\hat{\Gamma)})=q(\hat{\Gamma})\alpha$. Using Eq.\ref{Prob1} and Eq.\ref{Prob2}, the ratio of these two minimal path probabilities can be written as, 
\begin{equation}\label{SIproof2}
    \frac{\Pr(\Gamma)}{\hat{\Pr(\Gamma})}=\frac{q(\Gamma)q(\Theta_i)}{q(\hat{\Gamma})q(\Theta_j)}=\mathrm{e}^{S(\Gamma)}\left( \frac{p_i}{p_j}\right),
\end{equation} 
where we used the fact that $\mathrm{e}^{S(\Gamma)}=\frac{q(\Gamma)}{q(\hat{\Gamma})}$ and $\frac{q(\Theta_i)}{q(\Theta_j)}=\frac{p_i}{p_j}$. The latter equality results from Eq.\ref{lf}. As mentioned in the main text, at thermodynamic equilibrium, the steady-state probabilities follow the Boltzmann distribution such that $p_i/p_j=\mathrm{e}^{-(F_i-F_j)}=\mathrm{e}^{-\Delta F}$.  Also $S(\Gamma)=\Delta F$. Accordingly, Eq.\ref{SIproof2} reduces to Eq.\ref{SIproof1}.
\subsection*{Bijective function- Example}
\begin{figure}[h]
\includegraphics[width=0.5\textwidth]{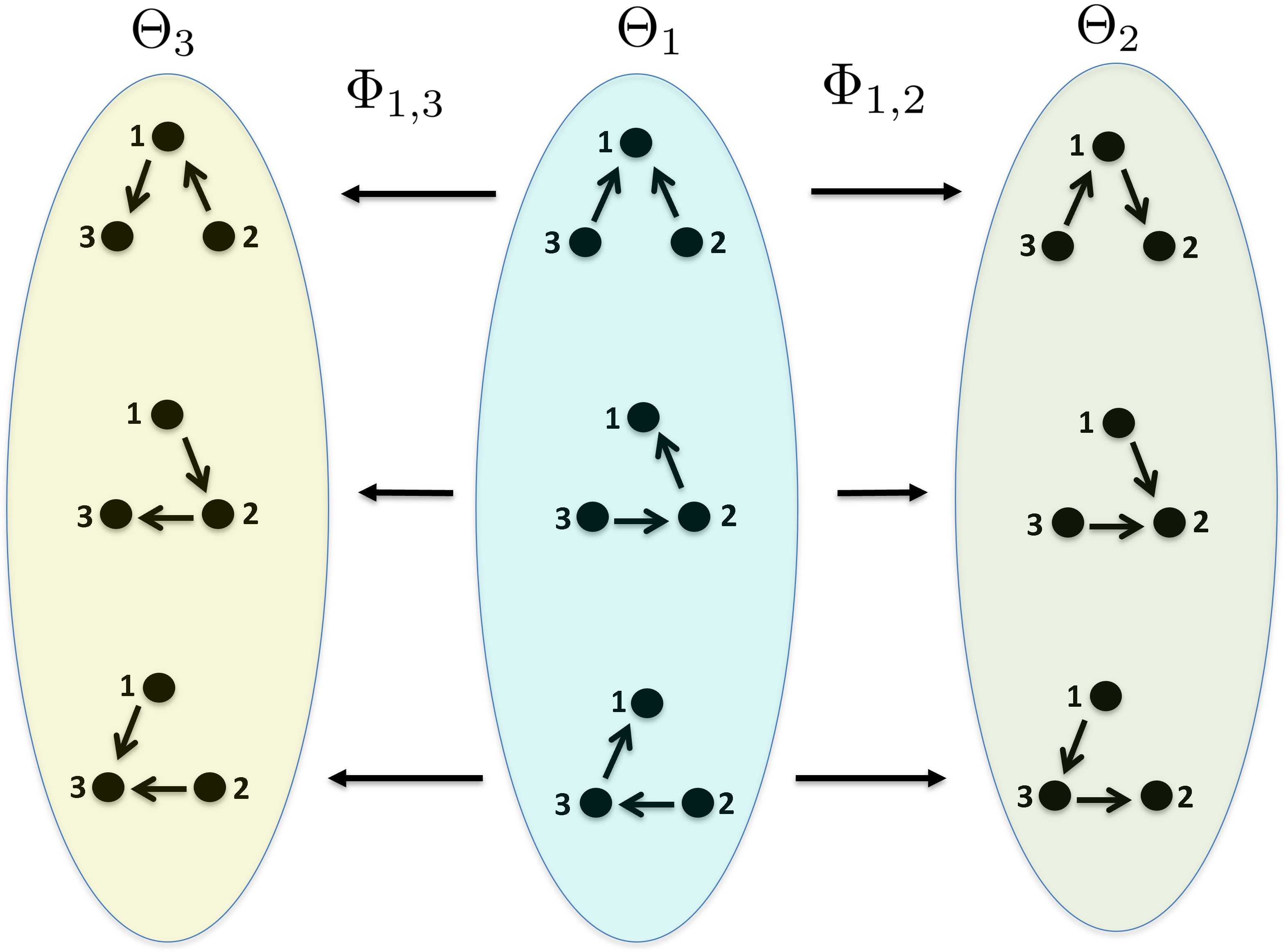}
\caption{The bijective function $\Phi_{i,j}\ : \Theta_i\rightarrow \Theta_j$ is described for the three-state graph in Fig.\ref{Fig1}. Here, $\Theta_i$ the set of all spanning trees rooted at vertex $i$. $\Theta_2$, for example, can be obtained from $\Theta_1$ using $\Phi_{1,2}$ which reverses the minimal paths from 2 to 1 on the spanning trees rooted at 1. The same reasoning applies to $\Theta_3$ and $\Phi_{1,3}$ as shown on the left.}
\label{Fig-bijection}
\end{figure}

We provide an example of the bijective function $\Phi_{i,j}\ : \Theta_i\rightarrow \Theta_j$ for the three-state graph in Fig.\ref{Fig1}. As shown in Fig.\ref{Fig-bijection}, the set of spanning trees rooted at vertex 3 and 2 can be obtained from the set of spanning trees rooted at 1 using the bijection defined previously.
\subsection*{Wilson's algorithm and the arboreal distribution}
In this section, based on Lawler's argument \cite{loopmeasures2018}, we provide more intuition through examples on how Wilson's algorithm can generate a random spanning tree rooted at the reference vertex according to the arboreal distribution. While these examples are not needed to follow the paper, we include them here for completeness. We refer to \citep{loopmeasures2018,lawler2013intersections,wilson1996generating} for a more detailed analysis of the Wilson algorithm and loop-erased random walks.

We first note that given a path generated by ${\cal L}(G)$ on $G$, the corresponding loop-erased path depends solely on the sequence of vertices visited along the path and is independent of the time spent at each vertex. Therefore, it is more convenient to work with the embedded discrete-time Markov chain when deriving the following results. Given a Markov process represented by $X(t)$, there exists an embedded discrete-time Markov chain denoted as $\tilde{X}(m)$, where $m$ belongs to the set of natural numbers $\mathbb{N} = \{0, 1, 2, \ldots\}$. The transition probabilities of the embedded discrete-time chain can be derived from the transition rates of the continuous-time Markov process. Specifically, the probability of transitioning from state $i$ to state $j$ in one step, denoted as $\Pr(\tilde{X}(m+1)=j| \tilde{X}(m)=i)$, is given by $\ell(i \to j)/\sum_{k \neq i} \ell(i \to k)$. We represent the  $n\times n$ transition probability matrix by ${\cal \tilde{L}}(G)$ whose $(j,i)$'th entry is ${\cal \tilde{L}}(G)_{ji}=\Pr(\tilde{X}(m+1)=j| \tilde{X}(m)=i)$. By construction, ${\cal \tilde{L}}(G)_{ii}=0$ for every $i$. Define the escape rate $\lambda_i$ for a vertex $i$ as the sum of transition rates from $i$ to all other vertices $k \neq i$, i.e., $\lambda_i=\sum_{k \neq i}\ell(i \to k)$. Let $I$ denote the $n\times n$ identity matrix and $\Lambda(G)$ represent the $n\times n$ diagonal matrix with entries given by $\Lambda_{ii}=\lambda_i$. We can rewrite the Laplacian matrix in terms of ${\cal \tilde{L}}(G)$, $I$ and $\Lambda(G)$ as follows, 
\begin{equation}\label{matrices}
{\cal L}(G)=({\cal \tilde{L}}(G)-I)\Lambda(G).
\end{equation}
Note that $\tilde{X}(m)$ also takes place on the same graph $G$, albeit with distinct edge labels, where transition rates are replaced by transition probabilities.
\subsubsection*{Example}
Let us provide the matrices defined above for the three-state graph in Fig.\ref{Fig1} and verify Eq.\ref{matrices}. 
\[
{\cal L}(G)=\begin{pmatrix}
-(\ell(1\ra 2)+\ell(1\ra3)) & \ell(2 \ra 1) & \ell(3\ra 1)\\
\ell(1 \ra 2)  & -(\ell(2\ra 1)+\ell(2\ra 3)) & \ell(3\ra2)\\
\ell(1\ra 3) & \ell(2\ra 3) & -(\ell(3\ra1)+\ell(3\ra 2))\\
\end{pmatrix},
\]
\[
\tilde{{\cal L}}(G)=\begin{pmatrix}
0 & \ell(2 \ra 1)/(\ell(2 \ra 1)+\ell(2\ra3)) & \ell(3\ra 1)/(\ell(3 \ra 1)+\ell(3\ra2))\\
\ell(1 \ra 2)/(\ell(1 \ra 2)+\ell(1\ra3))  & 0 & \ell(3\ra2)/(\ell(3 \ra 1)+\ell(3\ra2))\\
\ell(1\ra 3)/(\ell(1 \ra 2)+\ell(1\ra3)) & \ell(2\ra 3)/(\ell(2 \ra 1)+\ell(2\ra3)) &0\\
\end{pmatrix},
\]
\[
\Lambda(G)=\begin{pmatrix}
\ell(1 \ra 2)+\ell(1\ra3) & 0 & 0\\
0 & \ell(2 \ra 1)+\ell(2\ra3) & 0\\
0 & 0 & \ell(3 \ra 1)+\ell(3\ra2)\\
\end{pmatrix},
\]
\[
I=\begin{pmatrix}
1 & 0 & 0\\
0 & 1 & 0\\
0 & 0 & 1\\
\end{pmatrix}	.
\]

We can verify Eq.\ref{matrices} by inspection using these matrices. Below, we provide some useful theorems that may be helpful in understanding the connection between Wilson's algorithm and the arboreal distribution.  For brevity, we represent ${\cal L}(G)$, ${\cal \tilde{L}}(G)$, $\Lambda(G)$ as ${\cal L}$, ${\cal \tilde{L}}$, $\Lambda$, respectively.
\begin{theorem} \label{Thm2} For a subset $\Delta \subsetneq V=\{1,\dots,n\}$ of the state space, let $\tau^{\Delta}=\inf \{k\geq 1: \tilde{X}(k)\in \Delta\}$ be the first passage time to subset $\Delta$ and define $\mathcal{G}_\Delta(x,y)$ as the expected number of visits to $y$ starting from $x$ without entering $\Delta$ for $x,y \notin \Delta$, $\mathcal{G}_\Delta(x,y)=\sum_{i=0}^\infty \pr(\tilde{X}(i)=y, i<\tau^{\Delta}|\tilde{X}(0)=x)$. Define $\mathcal{G}_\Delta$ as a matrix whose entries are $\mathcal{G}_\Delta(x,y)$, then we have the following relation,
\begin{equation}\label{Thm2eq}
\mathcal{G}_\Delta=\left( I^\Delta -{\cal \tilde{L}}^\Delta\right)^{-1},
\end{equation}
and $\mathcal{G}_\Delta$ is called the Green function. 
\end{theorem}
A detailed derivation of this result can be found in textbooks on Markov processes \citep{lawler2013intersections}. A typical approach starts with the matrix expression below,
\[\mathcal{G}_\Delta=\sum\limits_{k=0}^\infty \left( {\cal \tilde{L}}^\Delta\right)^k,
\]
which yields $\left(I^\Delta- {\cal \tilde{L}}^\Delta \right) \mathcal{G}_\Delta=I^\Delta$. This implies that $\mathcal{G}_\Delta=\left( I^\Delta -{\cal \tilde{L}}^\Delta\right)^{-1}$.

Let $\tau^{x}=\inf \{k\geq 1: \tilde{X}(k)=x\}$ and define $f_x(\Delta)=\pr(\tau^x<\tau^\Delta)$ to be the probability that the random walk starting at $x$ returns to $x$ before entering $\Delta$, then,
\begin{equation}\label{g(x,x)}
\mathcal{G}_\Delta(x,x)=\sum_{k=0}^\infty f_x(\Delta)^k=\frac{1}{1-f_x(\Delta)}.
\end{equation}
This result is also a widely-known theorem in the theory of Markov chains. Due to its standard nature, we omit the proof here.

\subsection*{Establishing the connection}
We provide a brief overview of Lawler's argument \citep{kozdron2013determinants,loopmeasures2018}. Consider a path $P$ that starts from $u_1$ and terminates at the reference vertex 1. We want to determine the probability of obtaining a loop-erased trajectory, denoted as LE$(P) = [u_1, u_2, \dots, u_{m}]$, that exactly follows the specified pattern. To have LE$(P) = [u_1, u_2, \dots, u_{m}]$, the trajectory starting from $u_1$ can make multiple returns to $u_1$ (denoted as $r_1$) without entering $\Delta_1=\{1\}$. It then should move to the vertex $u_2$ with a probability given by $\ell(u_1 \to u_2)/\lambda_{u_1}$. Subsequently, the trajectory can make multiple returns to $u_2$ (denoted as $r_2$) without entering $\Delta_2=\{1,u_1\}$, and this process continues. Finally, it makes multiple returns to vertex $u_{m-1}$ (denoted as $r_{m-1}$) without entering $\Delta_{m-1}=\{1,u_1,\dots,u_{m-2}\}$, and then jumps from $u_{m-1}$ to $u_{m} = 1$ with a probability $\ell(u_{m-1} \to u_{m})/\lambda_{u_{m-1}}$. Using Eq.\ref{g(x,x)} and noting that $f_{u_i}(\Delta)$ is the return probability to vertex $u_i$ without entering $\Delta$, the probability of generating a loop-erased trajectory $\Gamma=\text{LE}(P) = [u_1, u_2, \dots, u_{m}]$ can be expressed as follows ,
\begin{equation}\label{LERWprob}
\begin{split}
\pr(\Gamma=[u_1,\dots,u_m])=\sum_{r_1=0,\dots,r_{m-1}=0}^\infty & f_{u_1}(\Delta_1)^{r_1}{\cal{\tilde{L}}}_{u_2u_1}\dots\\
\dots f_{u_{m-1}}(\Delta_{m-1})^{r_{m-1}}{\cal{\tilde{L}}}_{u_mu_{m-1}}.
\end{split}
\end{equation}
Using Eq.\ref{g(x,x)} and the definition of single-step jump probabilities, we can write Eq.\ref{LERWprob} as follows,
\begin{equation}\label{LERWprob2}
\begin{split}
\pr(\Gamma&=[u_1,\dots,u_m] )=\mathcal{G}_{\Delta_1}(u_1,u_1)\frac{\ell(u_1\to u_2)}{\lambda_{u_1}}\dots \\
&\dots \mathcal{G}_{\Delta_{m-1}}(u_{m-1},u_{m-1})\frac{\ell(u_{m-1}\to u_m)}{\lambda_{u_{m-1}}}.
\end{split}
\end{equation}
Let us use the three-vertex graph given in Fig. \ref{Fig1} as an example to show how these path probabilities are related to the arboreal distribution.

\subsection*{Example}
In this example, we find it convenient to label the three spanning trees rooted at 1 as $T_{\#1}$, $T_{\#2}$, and $T_{\#3}$ in left-to-right order in Fig.\ref{Fig1}. Stochastic trajectories that launch from vertex 2 and terminate at vertex 1 can give rise to the two possible distinct minimal paths after loop erasure: $[2\to1]$ and $[2\to3\to1]$. Prior to applying  Eq.\ref{LERWprob2} to calculate the corresponding probabilities, we should compute $\mathcal{G}_{\{1\}}$ and $\mathcal{G}_{\{1,2\}}$ using the Green function formula $\mathcal{G}_\Delta=\left( I^\Delta -\tilde{L}^\Delta\right)^{-1}$. The resulting values are as follows:
\renewcommand{\arraystretch}{2.5}
\begin{equation}
\mathcal{G}_{\{1\}}=\begin{pmatrix}
\dfrac{\lambda_2\lambda_3}{\sum_{T \in \Theta_1(G)} q(T)} & \dfrac{\ell(3\to2)\lambda_2}{\sum_{T \in \Theta_1(G)} q(T)} \\
\dfrac{\ell(2\to3)\lambda_3}{\sum_{T \in \Theta_1(G)} q(T)} & \dfrac{\lambda_2\lambda_3}{\sum_{T \in \Theta_1(G)} q(T)}\\
\end{pmatrix},
\end{equation}
and $\mathcal{G}_{\{1,2\}}=1$. Note that the original indices are preserved in these submatrices, e.g., $\mathcal{G}_{\{1\}}(2,2)=\frac{\lambda_2\lambda_3}{\sum_{T \in \Theta_1(G)} q(T)}.$ With all the necessary components at hand, we can now proceed to compute these minimal path probabilities using Eq.\ref{LERWprob2}, which yields,
\begin{equation}
\begin{split}
& \pr(\Gamma=[2\to1])=\mathcal{G}_{\{1\}}(2,2)\tilde{L}_{12}=\dfrac{\lambda_2\lambda_3}{\sum_{T \in \Theta_1(G)} q(T)}\dfrac{\ell(2\to1)}{\lambda_2}\\
&=\dfrac{\ell(2\to1)\ell(3\to2)+\ell(2\to1)\ell(3\to1)}{\sum_{T \in \Theta_1(G)q(T)}}\\
&=\pr_{\Theta_1(G)}(T_{\#1})+\pr_{\Theta_1(G)}(T_{\#2}),\\
\end{split}
\end{equation}
and
\begin{equation}
\begin{split}
&\pr(\Gamma=[2\to3\to1])=\mathcal{G}_{\{1\}}(2,2)\tilde{L}_{32}\mathcal{G}_{\{1,2\}}(3,3)\tilde{L}_{13}\\
=&\dfrac{\lambda_2\lambda_3}{\sum_{T \in \Theta_1(G)} q(T)}\dfrac{\ell(2\to3)}{\lambda_2}\dfrac{\ell(3\to1)}{\lambda_3}=\pr_{\Theta_1(G)}(T_{\#3}).\\
\end{split}
\end{equation}
Let us also consider the trajectories that launch from vertex 3 and terminate at the reference vertex. These trajectories create a probability distribution on the following minimal paths after loop erasure: $[3 \rightarrow 1]$ and $[3 \rightarrow 2 \rightarrow 1]$. Following essentially the same steps, we obtain the results below,
\begin{equation}
\begin{split}
& \pr(\Gamma=[3\to1])=\mathcal{G}_{\{1\}}(3,3)\tilde{L}_{13}=\dfrac{\lambda_2\lambda_3}{\sum_{T \in \Theta_1(G)} q(T)}\dfrac{\ell(3\to1)}{\lambda_3}\\
&=\dfrac{\ell(3\to1)\ell(2\to1)+\ell(3\to1)\ell(2\to3)}{\sum_{T \in \Theta_1(G)q(T)}}\\
&=\pr_{\Theta_1(G)}(T_{\#2})+\pr_{\Theta_1(G)}(T_{\#3}),\\
\end{split}
\end{equation}
and
\begin{equation}
\begin{split}
&\pr(\Gamma=[3\to2\to1])=\mathcal{G}_{\{1\}}(3,3)\tilde{L}_{23}\mathcal{G}_{\{1,3\}}(2,2)\tilde{L}_{12}\\
=&\dfrac{\lambda_2\lambda_3}{\sum_{T \in \Theta_1(G)} q(T)}\dfrac{\ell(3\to2)}{\lambda_3}\dfrac{\ell(2\to1)}{\lambda_2}=\pr_{\Theta_1(G)}(T_{\#1}),\\
\end{split}
\end{equation}
as expected. 
\end{document}